\documentclass[aip,jcp,amsmath,amssymb,floatfix,reprint,citeautoscript,noeprint]{revtex4-1}

\usepackage[utf8]{inputenc}
\usepackage{graphicx}
\usepackage{wrapfig}
\usepackage{bibunits}
\usepackage[colorlinks,allcolors=black,citecolor=blue,urlcolor=blue]{hyperref}
\usepackage[mathlines,displaymath]{lineno}
\usepackage{chemformula}
\usepackage{placeins}

\DeclareUnicodeCharacter{2212}{-}

\graphicspath{{figures/}}

\defaultbibliography{paper.bib}
\defaultbibliographystyle{aipnum4-1}

\begin{document}

\def\mytitle{
Machine learning potentials for complex aqueous systems made simple
}
\title{\mytitle}

\author{Christoph Schran}%
\email{cs2121@cam.ac.uk}
\affiliation{%
Yusuf Hamied Department of Chemistry, University of Cambridge, Lensfield Road, Cambridge, CB2 1EW, UK
}
\affiliation{%
Thomas Young Centre, London Centre for Nanotechnology, and Department of Physics and Astronomy, University College London, Gower Street, London, WC1E 6BT, United Kingdom
}%

\author{Fabian L. Thiemann}%
\thanks{These authors contributed equally.}
\affiliation{%
Yusuf Hamied Department of Chemistry, University of Cambridge, Lensfield Road, Cambridge, CB2 1EW, UK
}
\affiliation{%
Thomas Young Centre, London Centre for Nanotechnology, and Department of Physics and Astronomy, University College London, Gower Street, London, WC1E 6BT, United Kingdom
}%
\affiliation{%
Department of Chemical Engineering, Sargent Centre for Process Systems Engineering, Imperial College London,
South Kensington Campus, London SW7 2AZ, United Kingdom
}%

\author{Patrick Rowe}%
\thanks{These authors contributed equally.}
\affiliation{%
Yusuf Hamied Department of Chemistry, University of Cambridge, Lensfield Road, Cambridge, CB2 1EW, UK
}
\affiliation{%
Thomas Young Centre, London Centre for Nanotechnology, and Department of Physics and Astronomy, University College London, Gower Street, London, WC1E 6BT, United Kingdom
}%

\author{Erich A. Müller}%
\affiliation{%
Department of Chemical Engineering, Sargent Centre for Process Systems Engineering, Imperial College London,
South Kensington Campus, London SW7 2AZ, United Kingdom
}%

\author{Ondrej Marsalek}
\affiliation{%
Charles University, Faculty of Mathematics and Physics, Ke Karlovu 3, 121 16 Prague 2, Czech Republic
}%

\author{Angelos Michaelides}%
\email{am452@cam.ac.uk}
\affiliation{%
Yusuf Hamied Department of Chemistry, University of Cambridge, Lensfield Road, Cambridge, CB2 1EW, UK
}
\affiliation{%
Thomas Young Centre, London Centre for Nanotechnology, and Department of Physics and Astronomy, University College London, Gower Street, London, WC1E 6BT, United Kingdom
}%

%

%

\begin{abstract}
Simulation techniques based on accurate and efficient representations
of potential energy surfaces are urgently needed
for the understanding of complex aqueous systems such as solid-liquid interfaces.
Here, we present a machine learning framework that enables the efficient
development and validation of models for complex aqueous systems.
Instead of trying to deliver a globally-optimal machine learning potential,
we propose to develop models applicable to specific thermodynamic
state points in a simple and user-friendly process.
After an initial \textit{ab initio} simulation,
a machine learning potential is constructed
with minimum human effort through a data-driven active learning protocol.
Such models can afterwards be applied
in exhaustive simulations to provide reliable answers
for the scientific question at hand,
or systematically explore the thermal performance of \textit{ab initio} methods.
We showcase this methodology on a diverse set of
aqueous systems comprising bulk water with different ions
in solution, water on a titanium dioxide
surface, as well as water confined in nanotubes and between molybdenum disulfide sheets.
Highlighting the accuracy of our approach with respect to the underlying \textit{ab initio} reference,
the resulting models are evaluated in detail
with an automated validation protocol
that includes structural and dynamical properties and
the precision of the force prediction of the models.
Finally, we demonstrate the capabilities of our
approach for the description of water on the rutile
titanium dioxide (110) surface
to analyze the
structure and mobility of water on this surface.
Such machine learning models
provide a straightforward and uncomplicated but accurate extension of simulation time and
length scales for complex systems.
\end{abstract}

\date{This manuscript was compiled on \today}

{\maketitle}

\begin{bibunit}

There is a great need for a better understanding of complex
aqueous systems, in particular those involving solid-liquid interfaces,
to promote progress in fields as diverse
as heterogeneous catalysis, material design, biotechnology, and energy conversion or storage~\cite{Zaera2012/10.1021/cr2002068,Bjorneholm2016/10.1021/acs.chemrev.6b00045}.
For this purpose, atomistic insight provided by computational approaches is urgently required, but off-the-shelf simulation techniques come with important limitations.
\textit{Ab initio} based methods, such as \textit{ab initio} molecular dynamics (AIMD), struggle in terms of the accessible time and length scales, while traditional force field approaches are complicated to develop and often not accurate enough to provide reliable answers for complex interface problems.
In recent years, machine learning potentials (MLPs) have become a promising alternative,
bypassing expensive \textit{ab initio} calculations
and extending length and time scales in molecular simulations~\cite{Behler2016/10.1063/1.4966192,%
Butler2018/10.1038/s41586-018-0337-2,%
Deringer2019/10.1002/adma.201902765,%
Kang2020/10.1021/acs.accounts.0c00472,%
Behler2021/10.1021/acs.chemrev.0c00868%
}.
This is exemplified in studies on the understanding of the unique properties of water~\cite{Morawietz2016/10.1073/pnas.1602375113,Cheng2019/10.1073/pnas.1815117116,Gartner2020/10.1073/pnas.2015440117},
structural and electronic transitions in disordered silicon~\cite{Deringer2021/10.1038/s41586-020-03072-z},
and phase transitions of hybrid perovskites~\cite{Jinnouchi2019/10.1103/PhysRevLett.122.225701}
to name but a few.
The success of MLPs is grounded in a number of distinct approaches that
have been introduced over the years, notably using
artificial neural networks~\cite{%
Behler2007/10.1103/PhysRevLett.98.146401,%
Ghasemi2015/10.1103/PhysRevB.92.045131,%
Schuett2017/10.1038/ncomms13890,%
Zhang2018/10.1103/PhysRevLett.120.143001,%
Unke2019/10.1021/acs.jctc.9b00181%
},
or kernel based methods~\cite{%
Bartok2010/10.1103/PhysRevLett.104.136403,%
Rupp2012/10.1103/PhysRevLett.108.058301,%
Thompson2015/10.1016/j.jcp.2014.12.018,%
Shapeev2015/10.1137/15M1054183,%
Li2015/10.1103/PhysRevLett.114.096405,%
Chmiela2017/10.1126/sciadv.1603015%
}.

Despite compelling advances towards data-driven and automated techniques
mostly in the context of active learning~\cite{%
Gastegger2017/10.1039/c7sc02267k,%
Podryabinkin2017/10.1016/j.commatsci.2017.08.031,%
Zhang2019/10.1103/PhysRevMaterials.3.023804,%
Smith2018/10.1063/1.5023802,%
Deringer2018/10.1103/PhysRevLett.120.156001,%
Schutt2018/10.1063/1.5019779,%
Musil2018/10.1039/c7sc04665k,%
Schran2020/10.1021/acs.jctc.9b00805%
},
the construction of a successful model, in particular for complex systems,
remains a difficult task.
This becomes most apparent
when trying to achieve a high degree of
transferability or generality,
as for example recently shown in the development
of general purpose MLPs
for silicon~\cite{Bartok2018/10.1103/PhysRevX.8.041048},
carbon~\cite{Rowe2020/10.1063/5.0005084},
or phosphorous~\cite{Deringer2020/10.1038/s41467-020-19168-z}.
The examples cited and many similar ones reported in the literature can take years to develop as broad regions of phase space
have to be sampled by an appropriate balance of training points
to provide reliable predictions across the board.
As a consequence relatively few studies exist in which complex solid-liquid systems have been described with MLPs~\cite{%
Natarajan2016/10.1039/c6cp05711j,%
Hellstrom2019/10.1039/c8sc03033b,%
Andrade2020/10.1039/c9sc05116c,%
Ghorbanfekr2020/10.1021/acs.jpclett.0c01739,%
Artrith2019/10.1088/2515-7655/ab2060%
}.
These limitations have hampered progress in understanding solid-liquid interfaces,
where accurate MLPs are urgently needed and offer many opportunities
for deepening our understanding of processes like wetting, ice formation,
or liquid flow and friction under confinement.

However, the high degree of generality that most MLPs try to achieve is in practice not always needed to answer the scientific question at hand.
Often, it is sufficient to sample just a small region of configuration space under specific thermodynamic and boundary conditions, while reaching the time and length scales at appropriate
\textit{ab initio} accuracy represents the true challenge.
This is the main motivation behind the very promising on-the-fly learning techniques~\cite{%
Li2015/10.1103/PhysRevLett.114.096405,
Jinnouchi2020/10.1021/acs.jpclett.0c01061,
Vandermause2020/10.1038/s41524-020-0283-z},
(or surrogate models for global structure
optimizations~\cite{Bisbo2020/10.1103/PhysRevLett.124.086102})
that do not aim for a high degree of generality.
Yet, even these approaches have so far not had wide uptake for complex interfaces and they have been developed and validated on a system specific case by case basis. 
\begin{figure*}
\centering
\includegraphics[width=0.75\linewidth]{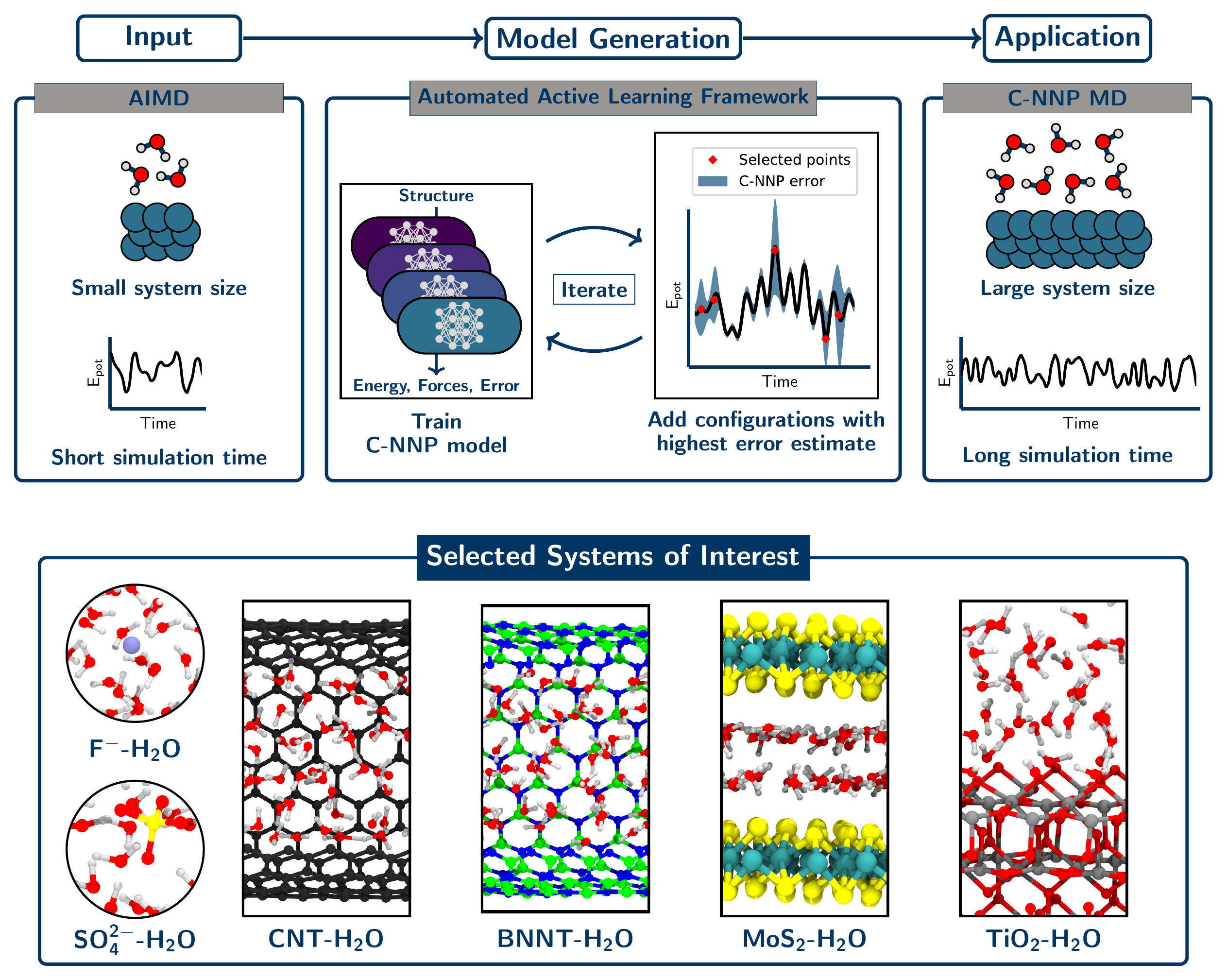}
\caption{\label{fig:workflow}
Schematic depiction of the rapid development process of
machine learning models with committee neural network potentials (C-NNPs).
The top panel depicts the workflow
used to generate a C-NNP model starting from
a single reference trajectory.
Using a small-scale \textit{ab initio} molecular dynamics (AIMD) simulation as input, the C-NNP
model is constructed in an active learning cycle that selects
the most important configurations for an improvement of the model.
This is achieved in an automated iterative process of first training the model and then screening of a large set of candidate configurations for structures with largest error estimate, which are added to the training set.
Subsequently, the C-NNP model can be applied to large-scale
simulations in order to provide insight into the system of interest.
The systems and potential energy curves schematically shown in the top panel are chosen for illustration purposes and do not reflect actual simulation data.
The bottom panel depicts representative sections of the simulation cells used for the
six aqueous systems chosen in this study for which we successfully
applied our machine learning protocol.
They are the fluoride ion in solution (\ch{F- -H2O}), the sulfate
ion in solution (\ch{SO4^{2-}-H2O}), water in carbon (\ch{CNT-H2O})
and hexagonal boron nitride nanotubes (\ch{BNNT-H2O}),
water under molybdenum disulfide confinement (\ch{MoS2-H2O}),
and water on a titatium dioxide interface (\ch{TiO2-H2O}).
}
\end{figure*}
Here, we present another approach to generate MLPs in a simple and fast fashion
that is particularly suited for complex systems.
To achieve this task, we make use of a 1) widely applicable, robust and scalable machine learning
model for the representation of the structure-energy relation,
2) a general strategy for the generation and selection of representative training data,
and finally 3) a comprehensive and automated validation procedure.
By design we concentrate the development on a specific thermodynamic condition.
This inherent loss of generality is counter-balanced
by the speed of the development as well as the local (as opposed to global) accuracy of the MLP.
The workflow to develop and apply such models follows broadly the following simple and computationally inexpensive steps.
The relevant thermodynamic condition is initially sampled with
a small-scale reference \textit{ab initio} simulation.
This trajectory is screened by a data-driven and automated
active learning procedure to construct the machine learning model.
The resulting model is then validated through a automated validation protocol and afterwards applied in large-scale simulations to answer the relevant scientific question.
We show how these models
can be developed with minimum human effort, while
retaining reliable predictions over long time scales for complex aqueous systems
at orders of magnitude lower computational cost than the original
\textit{ab initio} method.
This methodology is applied to six exemplary aqueous systems,
comprising the fluoride and sulfate ions in aqueous solution,
water in carbon and hexagonal boron nitride nanotubes,
water on a titanium dioxide surface,
and water under molybdenum disulfide confinement.
Due to the efficient nature of this approach, both from
a computational but also user perspective,
such readily developed models can afterwards
be applied in extensive molecular simulations to evaluate properties of interest.
We demonstrate this in the present case for the
description of water on the rutile titanium dioxide surface,
for which we investigate structural and dynamical properties
with extended molecular dynamics simulations.
We believe that the
change of paradigm
of generating a machine learning model in a ``cheap and simple'' process
as described here
will lead to an increase in the adoption of MLPs to simulate complex systems.
These concepts are also expected to be transferable to other machine learning
approaches that can be easily coupled to the open-source
active learning package, that we make freely accessible.
Having shown that this approach is able to correctly capture
the properties of various aqueous systems
under confinement and at interfaces,
we suggest that it outlines a straightforward
strategy for the uncomplicated but accurate investigation
of many technologically relevant systems.

\section*{Rapid development of committee neural network potentials}
\label{sec:c-nnp}
The implementation of our MLP framework
relies on committee neural network potentials (C-NNPs)~\cite{Schran2020/10.1063/5.0016004},
in our case built from Behler-Parrinello
NNPs~\cite{Behler2007/10.1103/PhysRevLett.98.146401,
Behler2017/10.1002/anie.201703114}.
While these models only consider the local environment of each atom explicitly within a finite range,
long-range contributions can in principle be incorporated in MLPs as demonstrated in the literature~\cite{%
Artrith2011/10.1103/PhysRevB.83.153101,
Grisafi2019/10.1063/1.5128375,
Ko2021/10.1038/s41467-020-20427-2,
Yue2021/10.1063/5.0031215%
}.
The main idea behind the current approach is the combination of multiple NNPs
in a ``committee model'', where the committee members are separately trained from independent random initial conditions
to a subset of the total training set.
The resulting committee model has multiple benefits over its individual members.
While the committee average, which is used as the prediction of the whole model,
has been shown to provide better performance than the individual NNPs,
the committee disagreement, which is defined as the standard deviation
between the committee member predictions, grants access to an estimate of the
error of the model.
If scaled by a constant obtained by comparison to the true validation error~\cite{Imbalzano2021/10.1063/5.0036522},
the committee disagreement provides an objective measure of the accuracy of the underlying model.
To construct a training set of such a model in an automated and data-driven way,
new configurations with the highest disagreement can be added to the training set.
This is an active learning strategy called query by committee and can be used to systematically improve a machine learning model, while
making efficient use of the limited data available which provides an important advantage for the current application.
Further details on the C-NNP methodology can be found in Ref.~\citenum{Schran2020/10.1063/5.0016004}.

These concepts can be used for the rapid development of MLPs
as described in the following
and schematically depicted in the top panel of Fig.~\ref{fig:workflow}.
Initially, a small-scale reference AIMD simulation is performed to sample the system of interest under the selected thermodynamic condition.
Small-scale refers here mostly to the envisaged time and length scales of the final application, while they might still be considered large from the \textit{ab initio} perspective.
In practice, AIMD trajectories with a length of 30\,ps have been sufficient for this purpose with system sizes that are large enough to cover all required local chemical environments.
Given this trajectory, which can also come from existing previous work, we then construct
the training set of the C-NNP model by selecting the most representative
configurations for the system and condition of interest in
an iterative active learning procedure.
In the beginning, a small set of 20 structures is randomly selected
from the reference trajectory in order to train an initial C-NNP model.
Next, query by committee is used to actively
select 20 configurations based on the highest committee disagreement in the atomic forces
from the set of candidate structures provided by the reference
trajectory.
These points are added to the training set and an improved
C-NNP model is trained to the extended training set.
No additional \textit{ab initio} reference calculations are required in this process,
as the potential energy and atomic forces are already available from the reference AIMD simulation.
Such iterations are repeated until convergence of the committee disagreement is observed, indicated by marginal improvements of the disagreement in subsequent iterations and no substantial difference in disagreement between the selected points and those already present in the training set.
This implies that
a sufficient variety of structures has been added to the training set to
yield an accurate and robust C-NNP model.
The uncomplicated construction of MLPs for new systems
requires a general set
of atomic descriptors.
In the case of the atom-centered symmetry functions~\cite{Behler2011/10.1063/1.3553717}, employed here, we
make that possible
by generating a systematic set of radial and angular functions.
This set consists of ten equidistantly shifted radial functions
with a fixed width and four angular functions
applied to each pair and triple of atoms, respectively.
In addition, we apply the same hyperparameters, such as
number of committee members, hidden layers and nodes, 
as well as neural network optimization parameters, to every system
to remove as much user input from our procedure as possible.
Thanks to the active learning procedure that adapts
the training set to the flexibility of a particular model,
we have in practice observed no limitations of such an application
of a general set of parameters and settings.
The remaining task of the user is to provide the reference
trajectory for the system of interest under the chosen simulation
conditions.
Given that input, a C-NNP model
can be obtained in practice without further adjustments
and in a short amount of time.

Once a new model is trained, it can be applied to
the system of interest close to
the same state point sampled by the original reference trajectory, reducing
the computational cost compared
to the \textit{ab initio} reference by orders of magnitude.
During such simulations, the committee disagreement of
the C-NNP model is a valuable tool to gauge
the validity of the model.
It provides an intrinsic error estimate
and thus can be monitored over the course of
the simulation and compared to the training process.
These concepts enable
the straightforward extension of
both time and length scales in molecular simulations.
Per design, the construction in the above fashion will be state dependent.
The heart of the matter is that this lack of generality is compensated
by the speed and simplicity in its fitting.

Besides this simple and robust framework
for the generation of MLPs for complex systems, special emphasis lies
on the selection of a suitable electronic structure reference method.
MLPs will always only be as good as their underlying reference method and a user
has to make a careful choice for each system of interest.
In this context, density functional theory (DFT) has become indispensable for
the investigation of aqueous systems, while careful benchmark studies~\cite{Grossman2003/10.1063/1.1630560,
Distasio2014/10.1063/1.4893377,
Forster-Tonigold2014/10.1063/1.4892400,
Gillan2016/10.1063/1.4944633,
Chen2017/10.1073/pnas.1712499114,
Marsalek2017/10.1021/acs.jpclett.7b00391,
Brandenburg2019/10.1063/1.5121370,
Schienbein2020/10.1002/anie.202009640}
can guide the selection of suitable functionals.
In addition, there has been promising steps to better understand remaining
limitations of existing functionals and provide potential solutions
in recent studies~\cite{Sharkas2020/10.1073/pnas.1921258117,
Wagle2021/10.1063/5.0041620,
Duignan2021/10.1021/acs.accounts.1c00107}.
Combined with the inexpensive yet reliable representation of interactions, as shown
in this work, this opens up the possibility for the uncomplicated but accurate
investigation of many technologically relevant systems.

Following our active learning workflow, we have obtained
C-NNP models for six different aqueous phase systems,
which are shown in the bottom panel of Fig.~\ref{fig:workflow}.
These diverse systems involve various types of interactions,
feature bulk, interface and confinement regions and
go up to a chemical composition of four elements.
In all cases, the training set of the model is exclusively
based on a set of structures generated by \textit{ab initio}
molecular dynamics simulations, as described
in detail in the Materials and Methods section.
Given that input, all six models have been generated
without further adjustments.
For all these systems, convergence was achieved with a compact training set of
roughly 300 structures,
highlighting the advantages of the active learning procedure.

\section*{Automated quality assessment of committee neural network potentials}
\label{sec:res} 
An automated training procedure calls for an efficient and robust validation protocol. 
Through extensive comparisons of our models to the underlying 
\textit{ab initio} reference trajectories we have identified a general set
of properties that serve to provide a stern test of our models.
These can be evaluated for any system of interest and provide a broad overview of the performance of a MLP, while further tests are included in the Supporting Information.
The selected properties exemplify the
performance of the models for structural and dynamical
properties as well as the precision of the force prediction.
Specifically, the performance for structural properties is assessed
by the match of the radial distribution functions (RDF)
for all involved species comparing the \textit{ab initio}
reference to the model prediction based on molecular dynamics
simulations.
All RDFs of a given system provide a comprehensive
summary of the structural arrangement of the
system of interest and are thus ideal to evaluate
the performance of the machine learning model
for thermodynamic properties.
Dynamical properties are validated by comparing the
species-resolved vibrational density of states (VDOS)
obtained with the model and the reference,
which gives a comprehensive overview of inter-
and intramolecular motions.
Finally, the force prediction of the model is validated by the
force root mean square error (RMSE) of a randomly selected
subset of structures from the \textit{ab initio}
reference simulation.
This quantity is chosen since the forces ultimately
drive the molecular dynamics simulations when using the model.
In order to make these properties comparable for
all systems, they are reduced into a score
by suitable difference measurements and subsequent
averaging over the involved species
as described in detail in the Supporting Information.
The entire testing protocol functions in an automated manner and efficiently provides a condensed summary of the accuracy of each model.

\begin{figure*}
\centering
\includegraphics[scale=1.0]{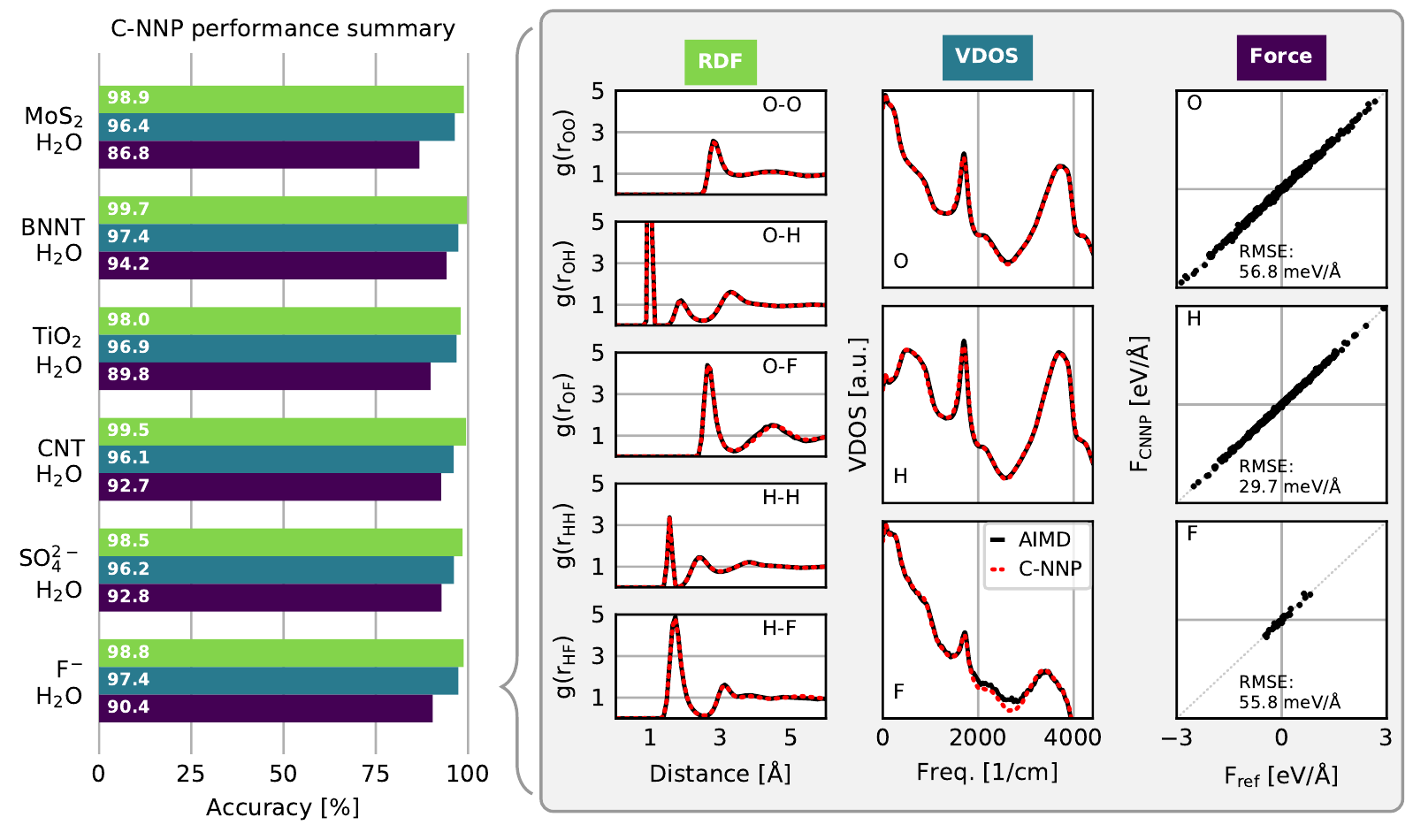}
\caption{\label{fig:score}
Performance assessment of the committee
neural network potentials (C-NNP) for six different aqueous systems.
The bar plot on the left features the summary of the accuracy
for the radial distribution functions (RDF), the
vibrational density of states (VDOS) and the force
predictions (Force) in percent for each system.
The plots on the right depict the species resolved
functions (RDF, VDOS (in logarithmic scale)) and the force correlation of the
C-NNP model with respect to the reference method
for the solvated fluoride ion (\ch{F- -H2O}) C-NNP model,
which are condensed into the three scores for the fluoride/water C-NNP model, shown on the left.
Details of the suitable difference measure and reduction for the three properties can be found in the Supporting Information.
}
\end{figure*}

The resulting summary of the quality assessment for all six studied systems is
shown in Fig.~\ref{fig:score}.
From this analysis it is clear that all models
reproduce the three selected
properties with rather high precision, where the RDF score ranges between
100 to 98\%, the VDOS score between
98 to 96\%, and the force score between
95 to 86\%.
To illustrate the meaning of these values, we
depict the individual functions (RDF and VDOS)
and the force correlation for the solvated fluoride
ion C-NNP model along with the total scores
in Fig.~\ref{fig:score}, while all other properties
for the remaining models are compiled in the Supporting Information.
This comparison shows that the selected
properties are reproduced with good
agreement to the \textit{ab initio} reference method
by our six C-NNP models.
In addition, all C-NNP results included in this
performance summary are based on substantially extended 
simulation times compared to the AIMD references as described in detail
in the Supporting Information.
This highlights the robust nature of our models, enabling
reliable predictions over long time scales.
We note that statistical fluctuations due to the more converged
nature of our C-NNP simulations accounts only for very minor changes of
the final property scores on the order of 0.5\%.

Besides the three sets of properties that we have quantitatively validated here, we have performed additional performance tests, in particular for the more complex systems of water confined in nanotubes and between \ch{MoS2} sheets as well as water on \ch{TiO2}.
These tests include the detailed analysis of the global structure of the solid and liquid subsystems, as encoded by the density profiles, the hydrogen bonding of water in the various systems, and the orientation of water with respect to the involved interfaces.
For all these tests, which are presented in detail in the Supporting Information, we observe good agreement between our C-NNP results and the AIMD reference simulations within the statistics of those shorter AIMD runs.
We are therefore confident that our performance overview, presented in Fig.~\ref{fig:score}, underlines the high quality of our C-NNP models.

The quality assessment for the six different systems included
in this work clearly highlights that our simple and straightforward
process to develop machine learning potentials
is able to provide robust and accurate models for the selected
thermodynamic condition.
Compared to the typical DFT setups employed here, the evaluation of the potential energy and atomic forces is usually four to five orders of magnitude faster with the C-NNP model.
As a consequence, all chosen systems could now be studied
in detail using exhaustive simulations that are accessible
with the developed models.
Given the focus on general properties in the testing protocol, we expect that it could prove useful for the development of potentials for various other solid-liquid systems
of technological and/or scientific interest.

\section*{Reaching longer length and time scales}
\label{sec:application}
Let us finally showcase the potential of the presented methodology
to extend the length and time scales of molecular simulations
and thus further the understanding of a system of interest.
For that purpose we investigate structural and dynamical properties
of water in contact with rutile \ch{TiO2}(110).
This system is of scientific and technological
importance due to the application of \ch{TiO2} for example in
photocatalysis or self-cleaning coatings and sensors.
In addition, it is an established prototypical oxide system
in surface science~\cite{Pang2013/10.1021/cr300409r}
and a rather controversial benchmark system
both for theory and experiment~\cite{Diebold2017/10.1063/1.4996116}.
For example, the extent of the mobility of water in the contact layers,
relevant e.g. for a detailed understanding of catalytic processes,
has been the focus of substantial research~\cite{%
Predota2007/10.1021/jp065165u,%
Liu2010/10.1103/PhysRevB.82.161415,%
Spencer2009/10.1021/jp8109918,%
English2012/10.1080/00268976.2012.683888,%
Agosta2017/10.1063/1.4991381%
}.

In order to shed light on these questions,
we have used the developed C-NNP model
to simulate rutile \ch{TiO2}(110)
in contact with water.
The model was constructed from a 30~ps AIMD simulation
at 300~K with the optB88-vdW functional~\cite{Klimes2010/10.1088/0953-8984/22/2/022201}, involving a
four O-Ti-O trilayer slab in contact with 80 water molecules,
forming a 1.5~nm water film on the surface.
After the development and benchmarking of
the C-NNP model as shown in the previous section,
we made a 2$\times$2 model of the interface (resulting in a \ch{TiO2/H2O} setup with 1728 atoms) and ran 5 ns of MD.
Reaching such length and time scales with AIMD simulations would represent an enormous computational burden, while they can be routinely performed with the C-NNP models.
Further details of these simulations can be
found in the Materials and Methods section.

\begin{figure}
\centering
\includegraphics[scale=1.0]{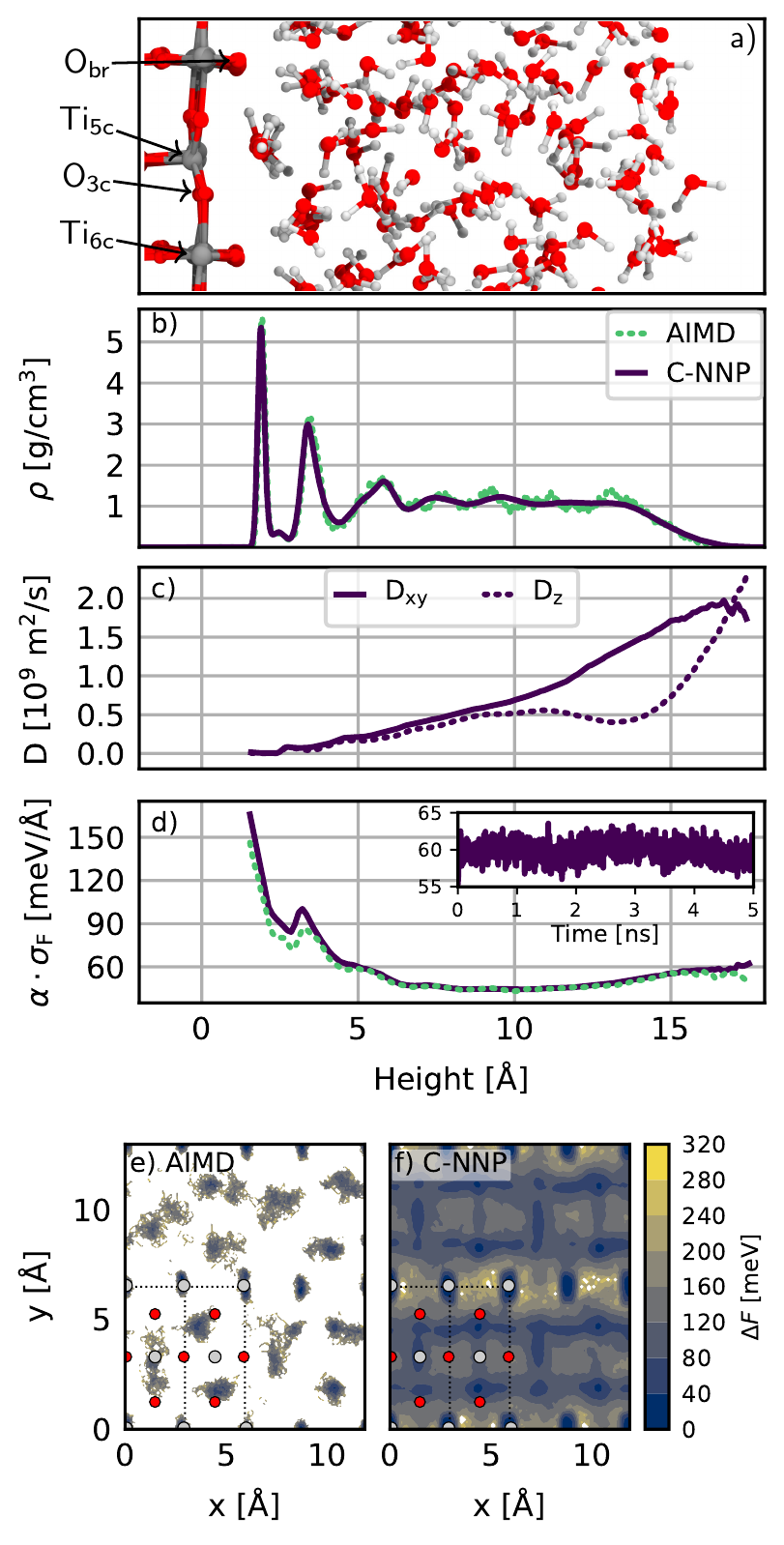}
\caption{\label{fig:tio2}
Properties of water on the rutile \ch{TiO2}(110) surface.
Panel a) depicts a representative section of the simulation
cell including the four distinct adsorption sites at the
interface (Ti$_\text{5c}$: fivefold coordinated titanium,
Ti$_\text{5c}$: sixfold coordinated titanium,
O$_\text{3c}$: threefold coordinated oxygen,
O$_\text{br}$: oxygen bridge site.
Panel b) features the mass density profile based on all water atoms,
panel c) the water diffusion constant separated into parallel (xy) and perpendicular (z) components,
and panel d) shows, as a function
of the distance from the surface, the C-NNP atomic force error estimate for structures from the C-NNP simulation and for all structures from the original AIMD simulation.
This error estimate is obtained as a direct product of the committee disagreement $\sigma_\text{F}$ and a scaling factor $\alpha$ to match the force RMSE of a validation set as proposed in Ref.~\citenum{Imbalzano2021/10.1063/5.0036522}.
The inset in panel~d) depicts the average atomic force error estimate of the water atoms as a function of the simulation time.
Panel e) and f) depict the free energy profile of the water adsorbed in the two contact layers from AIMD and C-NNP simulations, respectively.
Titanium atoms are shown in gray, oxygen atoms in red and hydrogen atoms in white.
}
\end{figure}

First, we analyze the water structuring by
looking at the density profile of water on the \ch{TiO2} surface
shown in panel~b) of Fig.~\ref{fig:tio2}.
The first important observation is the close
match between the crude density profile obtained
for our AIMD simulation and the statistically
converged profile obtained with the significantly
extended C-NNP simulation.
Overall, we observe a highly structured arrangement
of water in the first two layers, which
correspond to the water adsorbed on the 5-coordinated
titanium site for the first peak and the water
around the 6-coordinated titanium site for the second peak,
as shown in the snapshot in the top panel.
This density profile is substantially more structured
than for AIMD simulations of water
on rutile~\cite{Liu2010/10.1103/PhysRevB.82.161415}
with the PBE functional,
which highlights the complex dependence of
interfacial properties on the chosen functional.
Given the improved understanding of the dependence
of the properties of water on the DFT functional,
in particular regarding the inclusion of dispersion interactions~\cite{Gillan2016/10.1063/1.4944633},
we conclude that modern DFT approaches predict
a highly structured arrangement of water on
the rutile surface reaching up to about 1~nm
into the liquid.
This is also in good agreement with the density profiles
obtained from MLP simulations of water
on the anatase (101) surface~\cite{Andrade2020/10.1039/c9sc05116c}
that used the SCAN functional as the reference method.

In a next step, we evaluate the water diffusion coefficient
resolved by the distance from the \ch{TiO2} surface.
Specifically, we make use of the 
mean square displacement, which we spatially decompose
based on the position of each water molecule at zero delay~\cite{Pluharova2019/10.1021/acs.jctc.8b00111} ---
an approach made feasible by the extensive statistics
provided by our C-NNP model.
We then obtain a local estimate of the diffusion coefficient
by the well-known Einstein relation, which can be evaluated separately
for the parallel (xy) and perpendicular (z) directions with respect to the interface.
Panel~c) in Fig.~\ref{fig:tio2} depicts the resulting
water diffusion constant D$_\mathrm{xy}$ and D$_\mathrm{z}$
as a function of the distance from the \ch{TiO2} surface.
As anticipated from the very structured density profile,
the mobility close to the surface is reduced substantially,
where essentially no diffusion is observed in the strongly
adsorbed contact layer.
Beyond the first and second layer, the diffusivity in the xy-direction
increases steadily up to the water-vacuum interface.
At the same time, D$_\mathrm{z}$ features a plateau around 1\,nm from the surface and a substantial increase in diffusion towards the vapor interface.
Overall, this analysis highlights the strong influence of the
\ch{TiO2} interface on the water diffusion stretching more than
1\,nm into the liquid.

Next, we address the accuracy of our extended
C-NNP simulations by analyzing the intrinsic error
estimate of our model, given by the atomic force committee disagreement
$\sigma_\text{F}$.
Panel~d) in Fig.~\ref{fig:tio2} resolves this local error
estimate of the atomic force components for all water atoms as a function of the distance from the surface.
This error estimate is a product of the committee disagreement $\sigma_\text{F}$, as directly provided by our C-NNP simulations, and a scaling factor $\alpha$ determined to match the force RMSE of a validation set as proposed in Ref.~\citenum{Imbalzano2021/10.1063/5.0036522}.
In addition, we have evaluated this error estimate with our C-NNP model for all configurations of the original AIMD simulation to assess if the increased system size or C-NNP generated structures lead to higher errors.
Overall, we observe error estimates between 40 to 80~meV/\AA{} over the entire
water region with only slightly higher values
close to the \ch{TiO2} surface, indicating the
increased complexity of the involved interactions
in this inhomogeneous region.
Furthermore, the error (averaged over all water atoms) does not deteriorate over the course of the 5\,ns long trajectory, fluctuating around an average of 60~meV/\AA{}, as shown in the inset of panel~d).
Such atomic force errors are similar or even smaller than those reported for other developed MLPs e.g. for pure water~\cite{Morawietz2016/10.1073/pnas.1602375113,Cheng2019/10.1073/pnas.1815117116,Gartner2020/10.1073/pnas.2015440117,Schran2020/10.1063/5.0016004}.
At the same time, the error estimate obtained for the
AIMD configurations features essentially the same distance resolved profile,
which reveals that our C-NNP simulations are
able to conserve their predictive power, while
substantially extending both time and length scales
of the simulations.

Finally, we analyze the free energy profile of water adsorbed in the first two contact layers on the \ch{TiO2} surface, as depicted in panels~e) and f) in Fig.~\ref{fig:tio2} for the AIMD and C-NNP simulation, respectively.
From the direct comparison between the AIMD and C-NNP results it is clear that the limited statistics of the shorter AIMD simulation is insufficient to provide reliable insight into this property.
Only with the extensive sampling enabled by the C-NNP model, the free energy profile can be fully converged.
The C-NNP free energy profile clearly underlines the strong preference for water adsorption above the fivefold coordinated titanium sites in the first contact layer and the slightly weaker adsorption of water around the sixfold coordinated titanium and threefold coordinated oxygen sites in the second contact layer.
In between these adsorption sites, substantial free energy barriers are observed, highlighting the immobile nature of the two contact layers as also revealed by the analysis of the water diffusion.
In summary, the extensive simulations with our 
C-NNP model, obtained in a straightforward and efficient
workflow, provide detailed insight into the
properties of water on the rutile surface.
We observe a pronounced water layering effect with strong density fluctuations and clear evidence of a highly structured arrangement of water
in the first adsorption layers.
In addition, our analysis of the water dynamics reveals almost no water diffusion
close to the interface and a strong influence on the diffusion stretching more than 1\,nm into the liquid.
The treatment of a complex interfacial system such as this one requires an accurate description of the binding at the various adsorption sites as well as long-time sampling of the dynamics. 
The C-NNP model developed here delivers on both fronts, which highlights the potential of our approach to deepen understanding of technologically relevant solid-liquid systems.

\section*{Conclusion}
\label{sec:con}
In this work, we have presented a
machine learning framework that makes the generation of MLPs simple.
We have also showcased its versatility for
the description of a range of
complex aqueous systems.
Making use of committee neural network potentials,
we have shown how
MLPs can be obtained in a straightforward and robust process
from a single reference simulation.
By essentially removing the need to adjust any hyperparameters, a
new system of interest can be tackled
in a direct, data-driven way.
We have demonstrated the potential of this approach
employing it for six complex liquid and solid-liquid systems and have evaluated
the quality of the resulting models in detail
for various properties underlining the high accuracy
of our models.
This important final step is realized with an automated validation protocol that is fully integrated into
our framework.
These developments are directly accessible to the community
as they build exclusively on open-source solutions
and we make our underlying software package and all templates available.

In its spirit similar to on-the-fly learning techniques, we depart from the goal of a high degree of
transferability or generality to concentrate
exclusively on the thermodynamic condition relevant for the
chosen scientific question.
Under these constraints, we have shown how a
robust and accurate machine learning potential can be obtained
with limited user input in an uncomplicated process.
We note that we have also explored the application of our models
for elevated and lowered temperatures, which was possible without
problems in a $\pm$30\,K regime.
During such simulations, the intrinsic error estimate of our approach is a very useful tool to gauge the validity of the results.
Due to these promising signs, we plan to systematically validate the robustness of our models to venture beyond the chosen thermodynamic state point, for example to describe variations in pressure.
%

We note that we built our development on established components, such as an efficient ML structure-energy representation and active learning concepts.
At the same time, we expect that the concepts laid out in this work are transferable to other MLPs and active learning approaches, making it possible to achieve similar results.
The novelty of our work lies in a change in perspective, where relatively little effort is put into creating an intermolecular potential which, in spite of concentrating on a sub-section of phase space, is still robust and accurate enough to be used to describe the non-trivial behavior of complex molecular systems.
The importance clearly is not in the individual pieces but rather in the end-to-end framework and the broad range of applications made possible.
This work therefore enables simulations that were not possible a short time ago, pushing forward the straightforward and reliable application of MLPs.

Looking to the future, we
have limited ourselves to aqueous systems with four species in this work.
However, we are confident that systems with more element types can also be tackled
with our approach.
In addition, we have not explored reactive processes in this study. 
Since MLPs are able to describe bond breaking and bond making events by design, we again anticipate the straightforward application of our methodology to such situations.
This will be especially important to investigate interesting surface reaction phenomena, such as water dissociation on reactive surfaces or the recently reported reversible hydrolysis of zeolites in contact with water~\cite{Heard2019/10.1038/s41467-019-12752-y}.
Key to a successful application in such situations will be the sampling of the relevant reactive process by the initial AIMD simulation used as input to our active learning protocol.
Finally, we are currently exploring a training protocol in which structures are generated by classical molecular dynamics as input for our active learning protocol to minimize the
need for expensive \textit{ab initio} calculations.
In this approach the expensive quantum computing engine is only used to obtain the \textit{ab initio} potential energies and atomic forces for those configurations identified to be most important for the generation of the model.
Such an approach has potential for additional cost savings over the one presented here and does not require expertise in AIMD simulations, thus more readily opening up the approach to researchers from the classical force field community.
%

In our six showcase applications of complex aqueous systems we have developed models with different DFT functionals, all of which represent reasonable choices for the aqueous systems studied.
However, this illustrates a broader issue which is that there is currently no `perfect' DFT functional for water and complex aqueous interfaces.
We believe that the approach developed here could become a valuable tool in this long-standing quest to find suitable DFT functionals.
Since our procedures provide the ability to reveal the true converged thermal performance of any
given functional for a realistic system, at a modest cost, the systematic exploration of the performance of DFT methods for complex disordered systems becomes feasible.
This makes it possible to go beyond the usual energetic benchmarks of relatively small systems in the absence of temperature and thus facilitates direct comparison with experiment.
Due to the moderate size of our training sets, our framework is also expected to be
easily extendable to more expensive \textit{ab initio} methods, e.g. at the hybrid DFT level
or considering explicit electron correlation, thus making these methods available for the realistic simulation of complex interfacial systems.

Overall, the developments reported herein will enable
the investigation of complex aqueous processes
such as water structuring in contact with interfaces and
wetting or ice formation on surfaces in a straightforward
manner.
Although here we applied it to aqueous systems, we believe
that the methodology will also prove useful for
other materials and liquids in contact with solids, as well as
general solvation phenomena, enabling the fast
screening of different materials at \textit{ab initio} accuracy.
It will also be particularly useful for situations where long sampling is required as for the exploration of free energy surfaces, or calculations of dynamical properties, such as the friction or viscosity of liquids in contact with interfaces.
In summary, this work outlines a straightforward strategy for the uncomplicated yet accurate investigation of many technologically and scientifically relevant systems by molecular simulations.

\section*{Materials and Methods}
\label{sec:methods}
The introduced machine learning framework has been
implemented in the AML Python package, which
interleaves the required simulation packages
and data manipulation steps in a user friendly
environment.
The AML package is available free of charge at \url{https://github.com/MarsalekGroup/aml}
and enables the straightforward generation of
a C-NNP model given a reference
trajectory as input.
With this code all six C-NNP models were developed
for the various aqueous phase systems studied here.

NNP optimizations were performed with the
open-source n2p2 code~\cite{Singraber2019/10.1021/acs.jctc.8b01092}
using the optimization parameters and symmetry functions as
provided in the template file in the associated data repository
for this paper.
All additional information on the C-NNP fitting procedure
can be found in the SI, while all training input files, training sets
and parameters of the final models are publicly available
at \url{https://github.com/water-ice-group/simple-MLP}.

The reference AIMD simulations used as the starting point
for our C-NNP models employed
quite different DFT settings, while all having been
performed with the CP2K
simulation package~\cite{Kuhne2020/10.1063/5.0007045}.
We provide full detail about these reference simulations
in the SI, but the typical simulation setups consist
of 64 to 110 water molecules reaching simulation times between
30 to 130\,ps.

MD simulations using the C-NNP models were also performed with CP2K,
which features an open-source implementation of the C-NNP
methodology since release 8.1.
All C-NNP simulations for our validation protocol were propagated
for at least 0.5~ns to allow for the converged computation of
the RDF and VDOS.
Further details of the validation protocol and the associated simulations
can be found in the SI, while the validation can be performed
with the AML Python package.
The C-NNP simulations of the \ch{TiO2} water system
were propagated for 5~ns for a 2$\times$2$\times$1 supercell of the original AIMD setup,
resulting in total in 1728 atoms making up 320 water molecules on
four O-Ti-O trilayers in a 23.672, 25.988, 42.0\,\AA{} periodic box.
This simulation employed a molecular dynamics time step of 1~fs, while
using deuterium masses for the hydrogen atoms.
The temperature of 300~K was maintained with a canonical sampling through velocity rescaling thermostat~\cite{Bussi2007/10.1063/1.2408420}.
\begin{acknowledgments}
We would like to thank Christopher Penschke for providing the water \ch{TiO2} AIMD trajectory.
C.S. acknowledges partial financial support from the
\textit{Alexander von Humboldt-Stiftung}.
This work was partially supported by the OP RDE project (No. CZ.02.2.69/0.0/0.0/18\_070/0010462), International mobility of researchers at Charles University (MSCA- IF II).
We are grateful to the UK Materials and Molecular Modelling Hub for computational
resources, which is partially funded by EPSRC (EP/P020194/1 and EP/T022213/1).
We are also grateful for computational support from the UK national high performance computing service, ARCHER, for which access was obtained via the UKCP consortium, funded by EPSRC grant ref EP/P022561/1.
\end{acknowledgments}

%
%
%

%

\end{bibunit}

\clearpage

\setcounter{section}{0}
\setcounter{equation}{0}
\setcounter{figure}{0}
\setcounter{table}{0}
\setcounter{page}{1}

\renewcommand{\thesection}{S\arabic{section}}
\renewcommand{\theequation}{S\arabic{equation}}
\renewcommand{\thefigure}{S\arabic{figure}}
\renewcommand{\thepage}{S\arabic{page}}
\renewcommand{\citenumfont}[1]{S#1}
\renewcommand{\bibnumfmt}[1]{$^{\rm{S#1}}$}

\title{Supporting information for: \mytitle}
{\maketitle}

\onecolumngrid

\begin{bibunit}

\section*{Quality assessment}
\label{sec:scoring}
In order to validate the six models
developed by our rapid machine learning framework,
we have established a validation protocol
that makes it possible to compare the performance
of the various models for different systems
in a direct and condensed manner.
For that purpose we have selected three main categories
for structural and dynamical properties as well as
the precision of the force prediction.
The main idea behind this validation protocol
is that it probes the performance of a given
model for the thermodynamic condition it
is developed for.
Thus, all three categories are compared
directly against the AIMD simulation that
was used as the starting point of the development of
the models.
This enables the straight-forward evaluation of the
models without the need for additional benchmark
simulations with the expensive reference method.

Structural properties for complex liquid-solid
systems are directly probed by the
radial distribution functions (RDFs) of the various species,
which provide detailed insight into the two-component
structural arrangement.
For the scoring of the models, we compute all RDFs
for a given system, both for the AIMD reference simulation
and for independent C-NNP simulations using the developed
model.
For an $N$ component system this results in $\binom{N+1}{2}$ RDFs
which can be directly compared between the AIMD and
C-NNP results by using a suitable norm $d^\text{RDF}$
\begin{align}
    d^\text{RDF} = 1 - \frac{\int_{0}^{+\infty} \left| g^\text{AIMD}(r)-g^\text{C-NNP}(r)\right|\text{d}r}
             {\int_{0}^{+\infty}g^\text{AIMD}(r) \text{d}r + \int_{0}^{+\infty}g^\text{C-NNP}(r)\text{d}r}
\end{align}
which provides a measure of the similarity of two RDFs
ranging from 0 (for most different) to 1 (for identical).
Averaging over all $\binom{N+1}{2}$ norms $d^\text{RDF}$ finally
yields a single number that can be converted into percent
to provide a condensed score for the performance
of the C-NNP model for structural properties.

Dynamical properties are directly encoded by the
vibrational density of states (VDOS), which is obtained
by Fourier transform of the velocity autocorrelation function.
The VDOS can be computed separately for all components in a
system of interest and thus provides detailed insight
into the dynamical properties of the system,
probing vastly different processes
over a broad frequency range.
For an $N$ component system $N$ species-resolved VDOS
are computed, both for the AIMD and C-NNP simulations.
Using a similar norm $d^\text{VDOS}$ as for the RDFs
\begin{align}
    d^\text{VDOS} = 1 - \frac{\int_{0}^{+\infty} \left| f^\text{AIMD}(\nu)-f^\text{C-NNP}(\nu)\right|\text{d}\nu}
             {\int_{0}^{+\infty}f^\text{AIMD}(\nu) \text{d}\nu + \int_{0}^{+\infty}f^\text{C-NNP}(\nu)\text{d}\nu}
\end{align}
the AIMD and C-NNP results can be condensed into a measure
of the similarity between the different functions, which
after averaging over the different species and conversion
into percent provides the score of a C-NNP model
for dynamical properties.

Finally, the precision of the C-NNP model for the prediction
of the forces is evaluated to provide another property score.
The forces are what drives the dynamics of the system of
interest and are thus of fundamental importance for an
accurate description.
We generated a test set for the evaluation of the force
performance, by selecting a large subset of 1000 structures
and associated forces from the original AIMD simulation.
The root mean square error (RMSE), calculated separately
for the $N$ species in each system
\begin{align}
    F^\text{RMSE} = \sqrt{\frac{\sum_{i=1}^{3M} \left(F^\text{AIMD}_{i} - F^\text{C-NNP}_{i}\right)^2}{3M}}
\end{align}
is used as a suitable measure for the force prediction.
Since the magnitude of the forces can fluctuate strongly
for different systems, but also within a given system (solid compared to
liquid atoms), the RMSE is put into relation of the average force
fluctuation of a given species
\begin{align}
    F^\text{RMS} = \sqrt{\frac{\sum_{i=1}^{3M} \left(F^\text{AIMD}_{i}\right)^2}{3M}}.
\end{align}
The resulting $N$ scaled force errors $F^\text{Force} = \frac{F^\text{RMSE}}{F^\text{RMS}}$
are averaged and converted into percent to provide the
force score of the C-NNP model.

\subsection*{Properties evaluated for Quality Assessment}
\label{subsec:prop}

\begin{figure}
\centering
\includegraphics[scale=1.0]{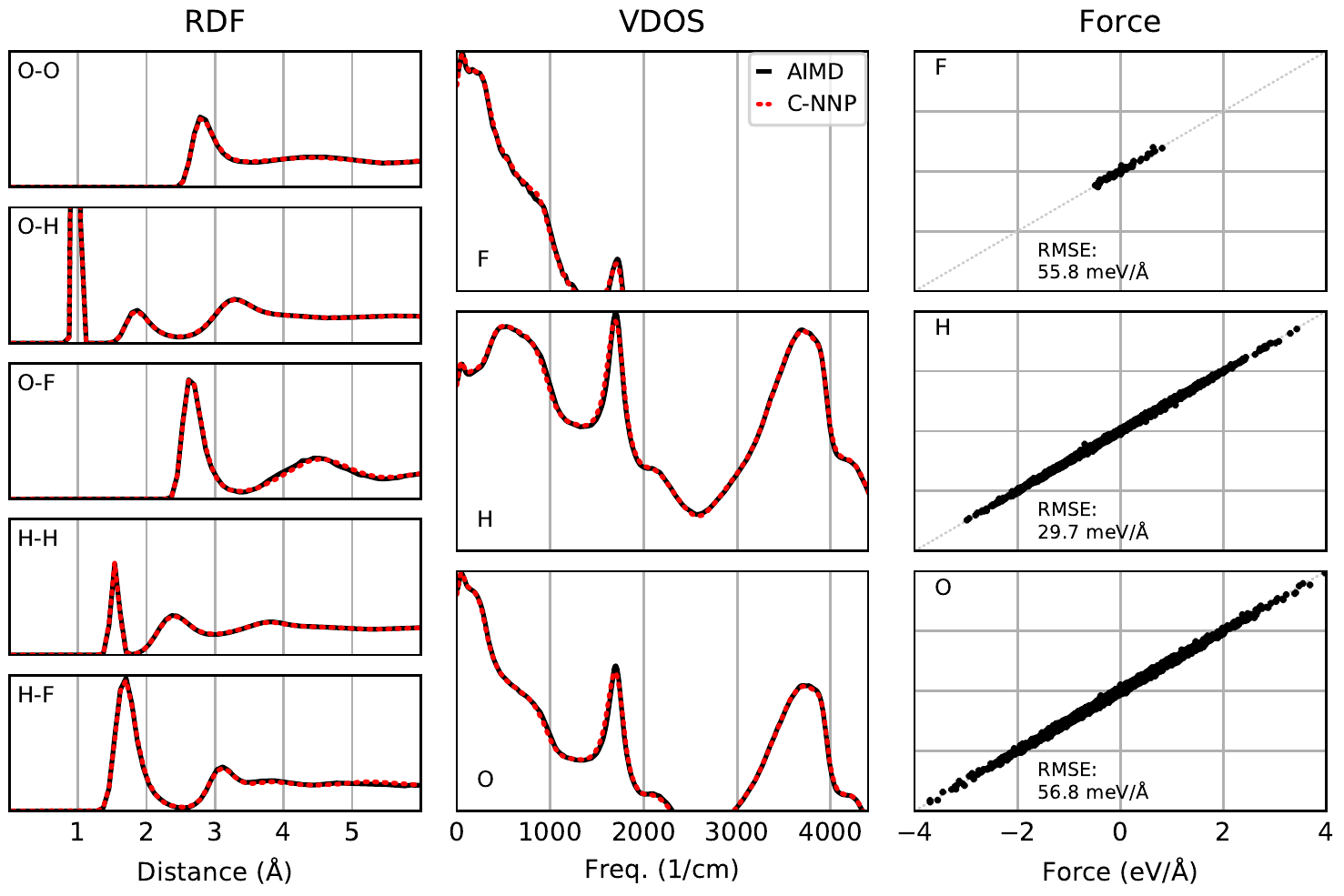}
\caption{\label{fig:score-f-h2o}
Assessment of the precision of the fluoride-water C-NNP model
for structural and dynamical properties as well as the force prediction.
The radial distribution functions (RDF) of all pairs of species
are shown in the left panels comparing the AIMD and C-NNP results.
The vibrational density of states (VDOS) of all species
are shown in the middle panels comparing the AIMD and C-NNP results.
The force correlation between AIMD and C-NNP forces of all species
are shown in the right panels.
}
\end{figure}

\begin{figure}
\centering
\includegraphics[scale=1.0]{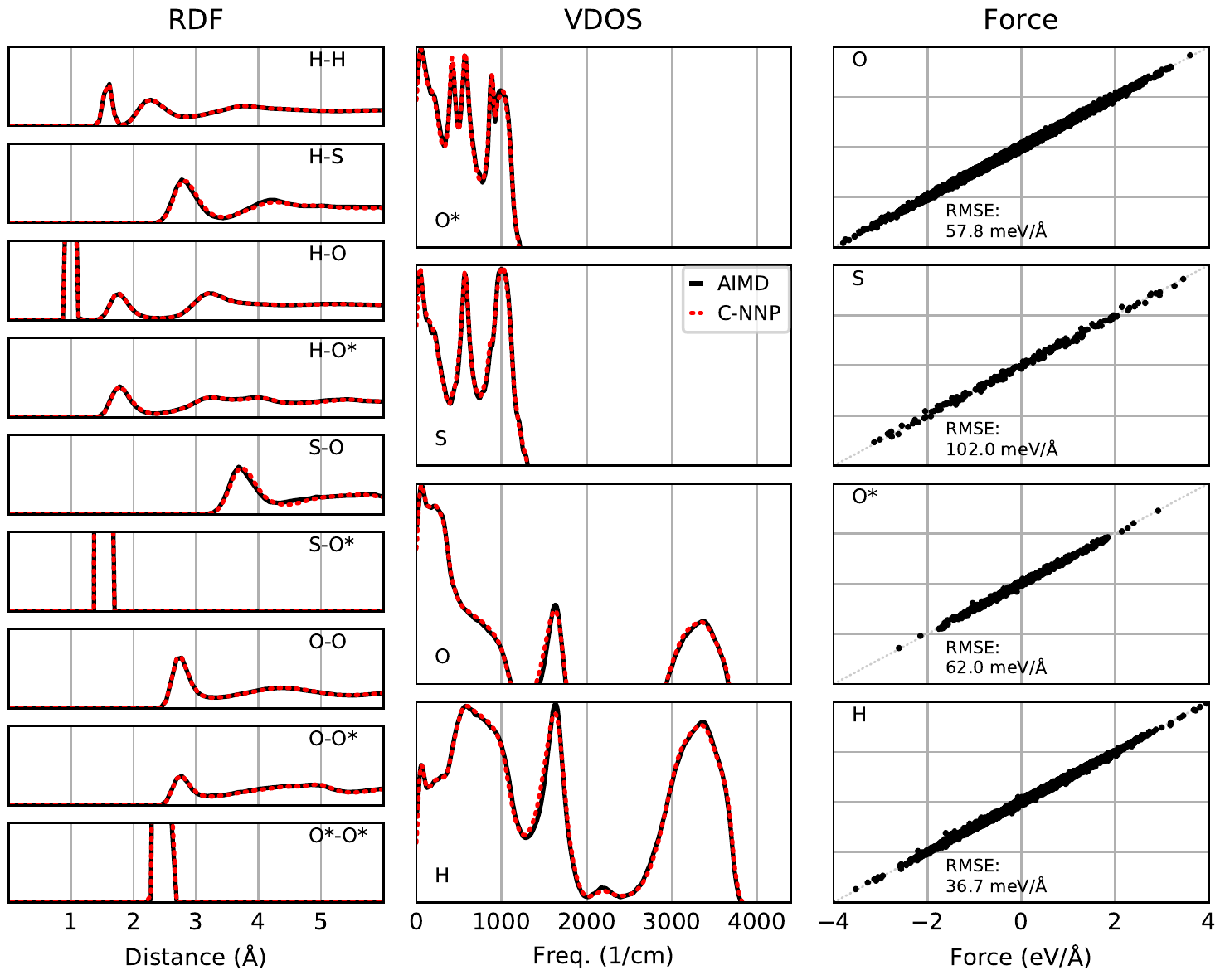}
\caption{\label{fig:score-so4-h2o}
Assessment of the precision of the sulfate-water C-NNP model
for structural and dynamical properties as well as the force prediction.
The radial distribution functions (RDF) of all pairs of species
are shown in the left panels comparing the AIMD and C-NNP results.
The vibrational density of states (VDOS) of all species
are shown in the middle panels comparing the AIMD and C-NNP results.
The force correlation between AIMD and C-NNP forces of all species
are shown in the right panels.
}
\end{figure}

\begin{figure}
\centering
\includegraphics[scale=1.0]{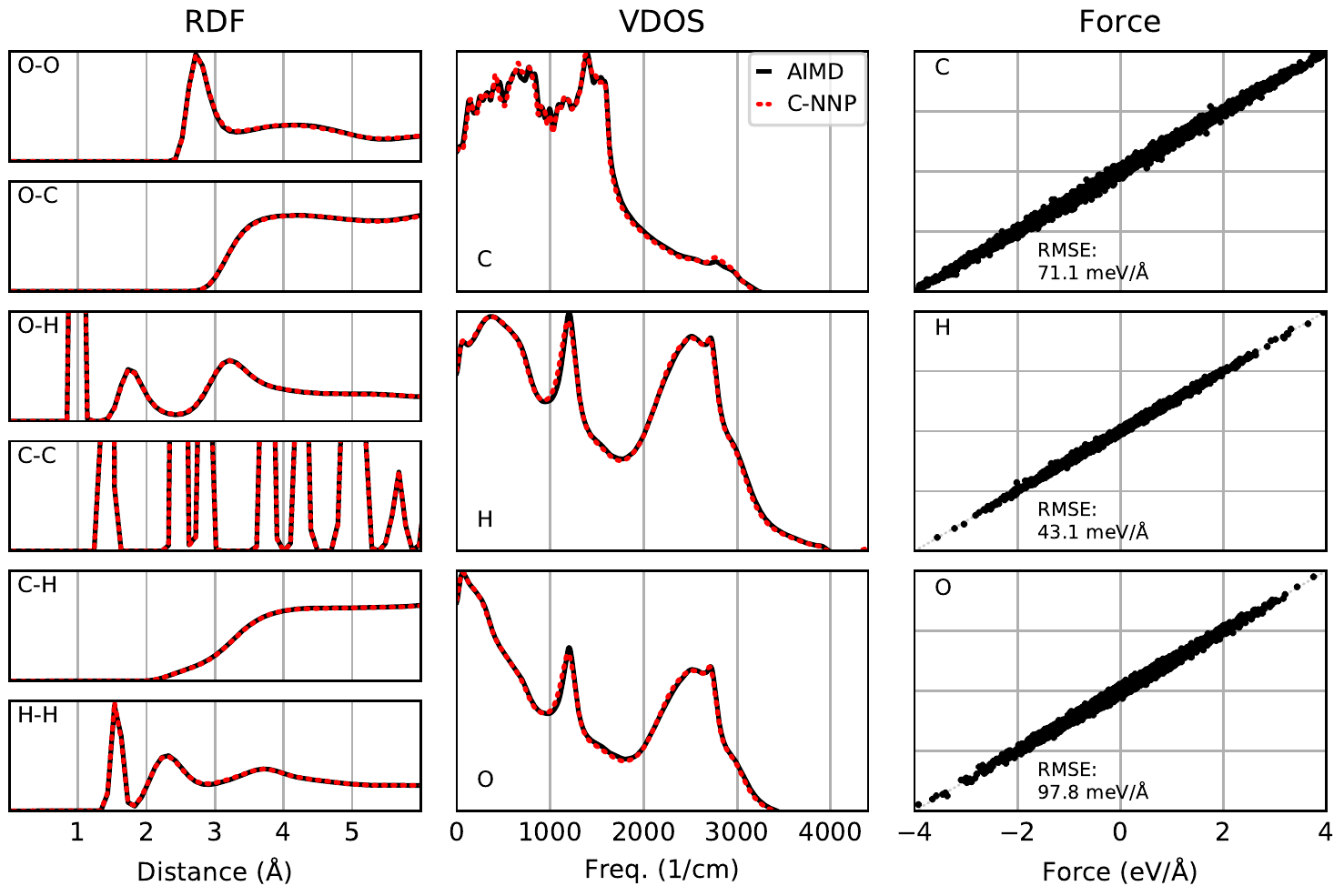}
\caption{\label{fig:score-cnt-h2o}
Assessment of the precision of the carbon nanotube-water C-NNP model
for structural and dynamical properties as well as the force prediction.
The radial distribution functions (RDF) of all pairs of species
are shown in the left panels comparing the AIMD and C-NNP results.
The vibrational density of states (VDOS) of all species
are shown in the middle panels comparing the AIMD and C-NNP results.
The force correlation between AIMD and C-NNP forces of all species
are shown in the right panels.
}
\end{figure}

\begin{figure}
\centering
\includegraphics[scale=1.0]{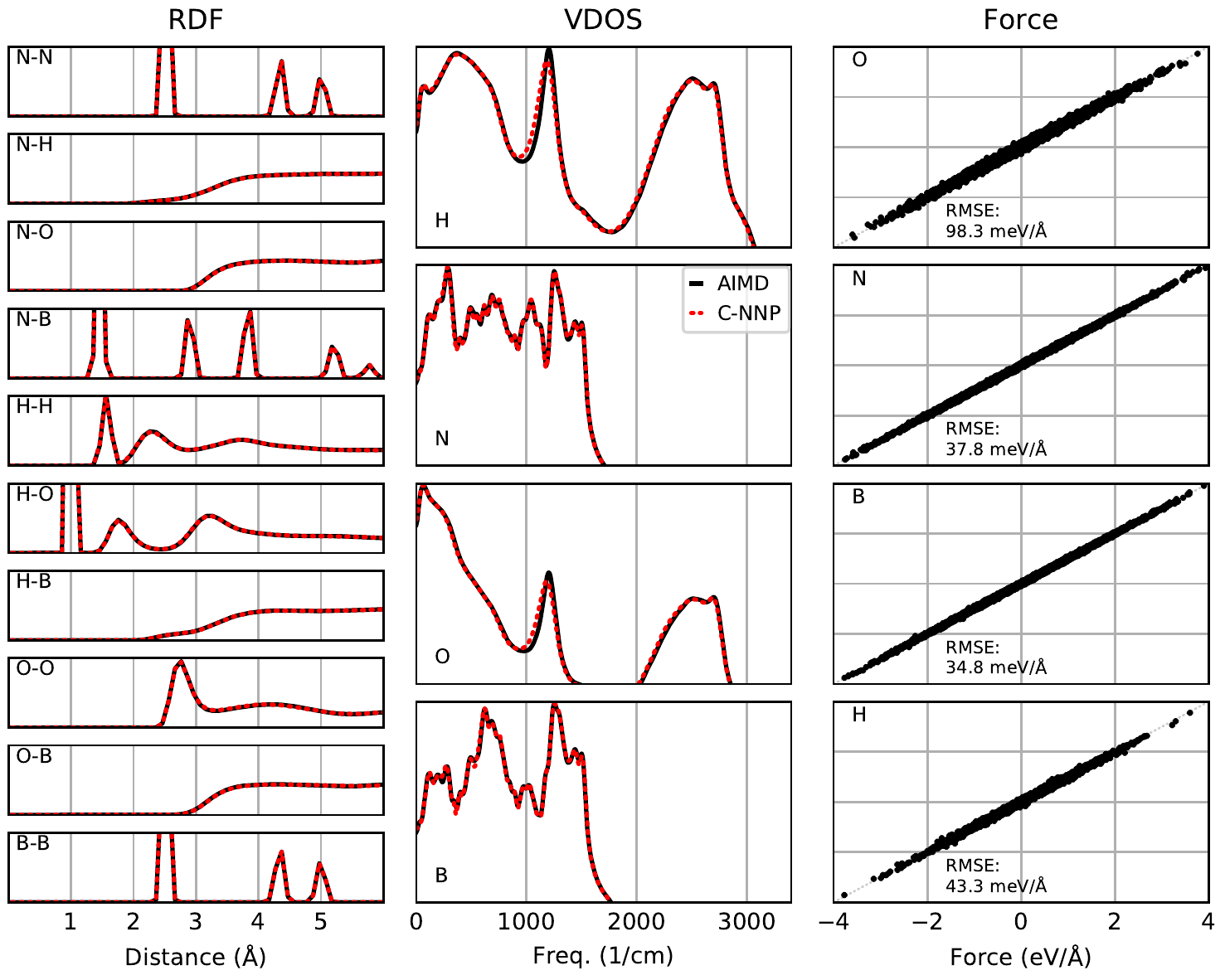}
\caption{\label{fig:score-hbn-h2o}
Assessment of the precision of the hexagonal boron nitride nanotube-water C-NNP model
for structural and dynamical properties as well as the force prediction.
The radial distribution functions (RDF) of all pairs of species
are shown in the left panels comparing the AIMD and C-NNP results.
The vibrational density of states (VDOS) of all species
are shown in the middle panels comparing the AIMD and C-NNP results.
The force correlation between AIMD and C-NNP forces of all species
are shown in the right panels.
}
\end{figure}

\begin{figure}
\centering
\includegraphics[scale=1.0]{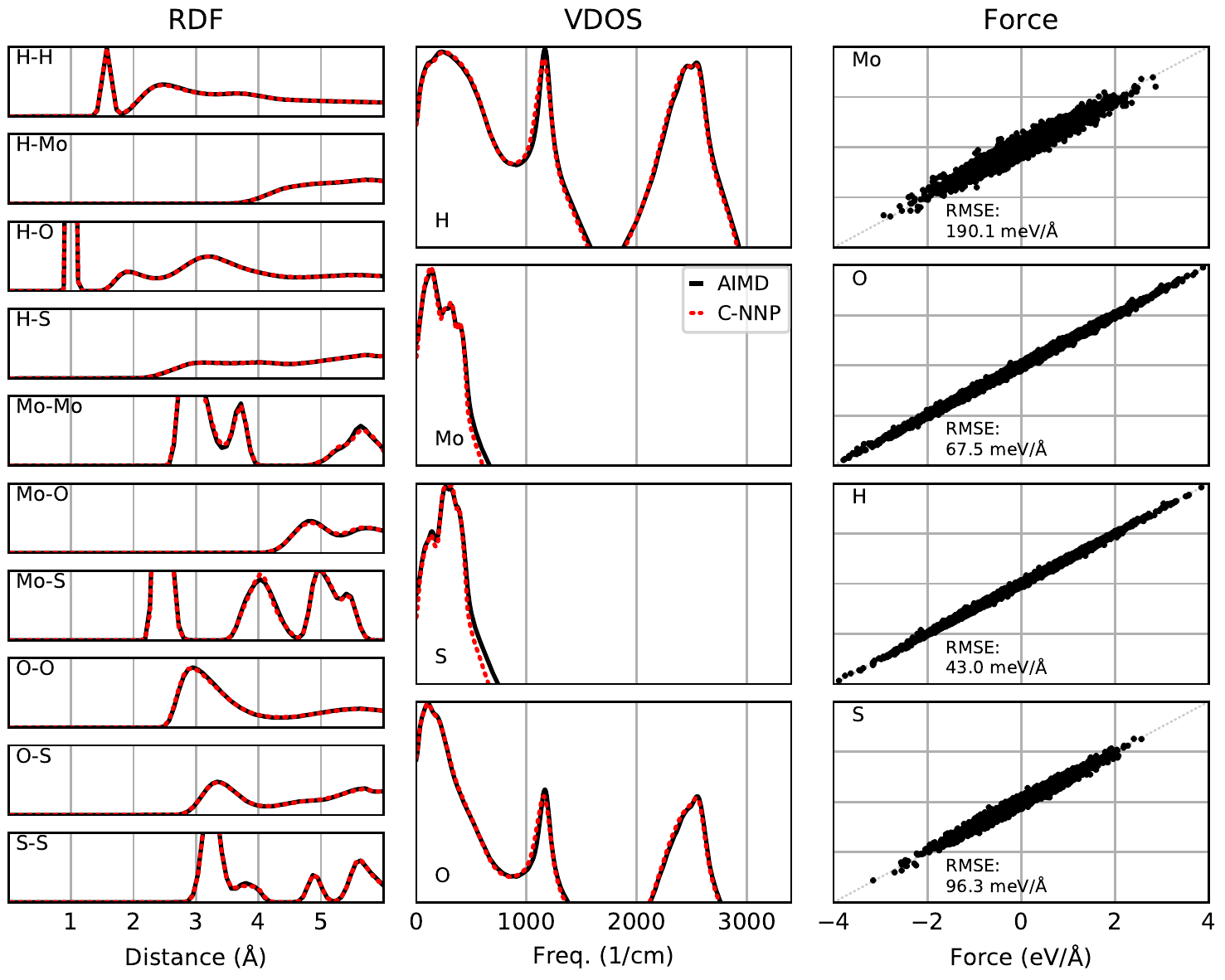}
\caption{\label{fig:score-mos2-h2o}
Assessment of the precision of the molybdenum disulfide-water C-NNP model
for structural and dynamical properties as well as the force prediction.
The radial distribution functions (RDF) of all pairs of species
are shown in the left panels comparing the AIMD and C-NNP results.
The vibrational density of states (VDOS) of all species
are shown in the middle panels comparing the AIMD and C-NNP results.
The force correlation between AIMD and C-NNP forces of all species
are shown in the right panels.
}
\end{figure}

\begin{figure}
\centering
\includegraphics[scale=1.0]{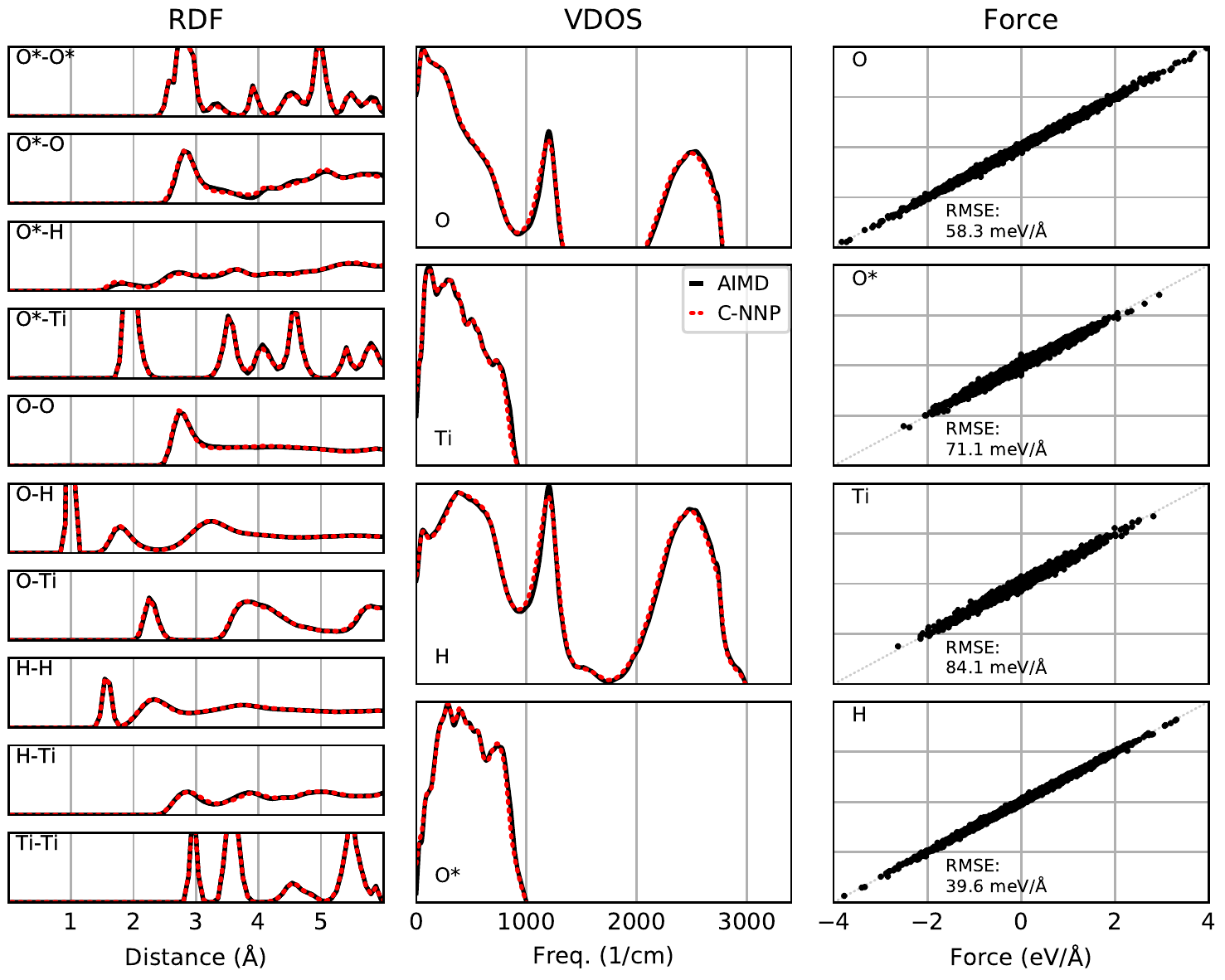}
\caption{\label{fig:score-tio2-h2o}
Assessment of the precision of the titanium dioxide-water C-NNP model
for structural and dynamical properties as well as the force prediction.
The radial distribution functions (RDF) of all pairs of species
are shown in the left panels comparing the AIMD and C-NNP results.
The vibrational density of states (VDOS) of all species
are shown in the middle panels comparing the AIMD and C-NNP results.
The force correlation between AIMD and C-NNP forces of all species
are shown in the right panels.
}
\end{figure}

All individual properties evaluated for the validation protocol
comprising the final RDF, VDOS, and force score, as presented in the main text,
are shown in full detail for all six C-NNP models
in Fig.~\ref{fig:score-f-h2o} to Fig.~\ref{fig:score-tio2-h2o}.
Overall, essentially perfect agreement between
the AIMD and C-NNP properties is observed for
all six systems.

\begin{figure}
\centering
\includegraphics[scale=1.0]{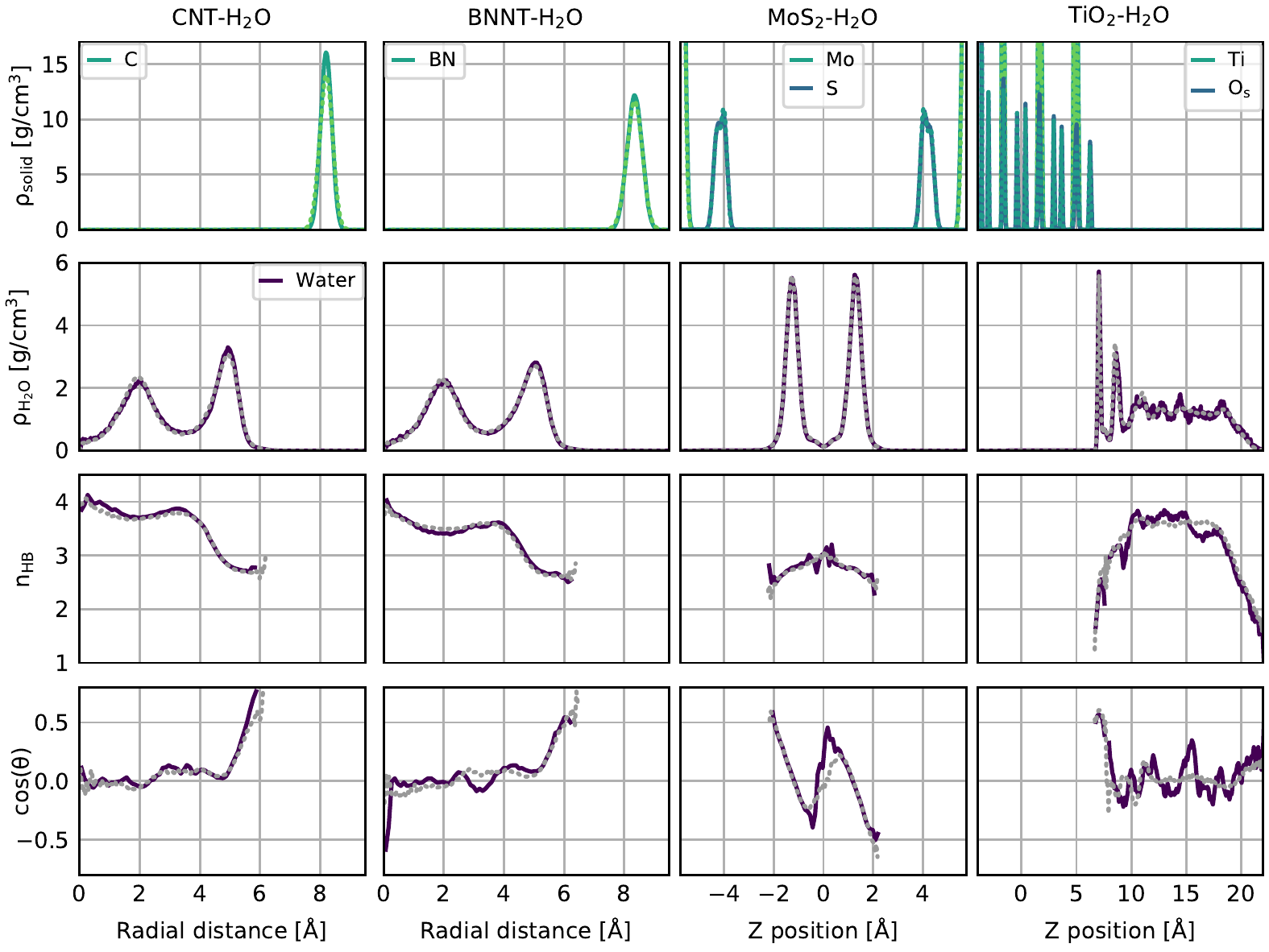}
\caption{\label{fig:other-prop}
Validation of system properties for four complex systems involving water under confinement or at interfaces.
The first, second, third, and forth column compile results for water in a carbon nanotube (\ch{CNT-H2O}), water in a hexagonal boron nitride nanotube (\ch{BNNT-H2O}), water confined by single layer molybdenum disulfide (\ch{MoS2-H2O}), and water at the rutile titanium dioxide surface (\ch{TiO2-H2O}), respectively.
The top row shows the density profiles of the involved solid subsystem and the second row the corresponding density of the water as a function of the radius for the two nanotubes and Z position for the others.
The third row depicts the number of hydrogen bonds along the water density profile
for the four systems, while the bottom row features the orientation of water with respect to the involved interface along the water density profile.
The AIMD reference results are shown with solid lines, while the C-NNP predictions are included as dotted lines.
}
\end{figure}

In addition, we have evaluated other properties
for the systems of water under confinement or at interfaces
in order to validate our C-NNP models in more detail.
These properties are the density profiles, number of hydrogen bonds, and water orientation
with respect to the interfaces for the \ch{CNT-H2O}, \ch{BNNT-H2O}, \ch{MoS-H2O},
and \ch{TiO2-H2O} systems.
They are all depict in Fig.~\ref{fig:other-prop} where overall substantial agreement
between the C-NNP prediction and the much shorter AIMD reference simulations is observed.
In particular, the density profile of the liquid and solid subsystems
as shown in the upper two columns highlight the different nature of
confinement with distinct density modulations that are all well reproduced
by our C-NNP models.
In addition, the number of hydrogen bonds along the water density
profile is in substantial agreement between AIMD reference and C-NNP prediction
for all four systems.
Finally, the orientation of water with respect to the involved interfaces,
as encoded by the cosine of the angle between the water dipole vector
and the normal of the interface, is also well reproduced by our C-NNP models.

\section*{Computational Details}
\label{sec:comp-det}
\subsection*{C-NNP models}
\label{subsec:c-nnp}

All C-NNP models for the six selected systems
were trained with the active learning workflow implemented
in the AML Python package, available
at \url{https://github.com/MarsalekGroup/aml}.
Using an \textit{ab initio} trajectory as input, we construct
a C-NNP model and its associated training set in an
active learning protocol.
20 randomly selected structures from the trajectory are used
to generate an initial C-NNP model.
Next, the model is improved by iteratively adding
structures with largest mean force committee disagreement to
the training set, which is continued until
convergence of the committee disagreement is observed.
We performed 15 such active learning steps for all
systems studied here, identifying 20 new structures
for the training set in every step, but keeping out previously
selected structures, which results in a
total training set size of about 300 structures.
Within every active learning step, the committee disagreement
of 2000 randomly selected structures from the AIMD reference
trajectory is evaluated and the 20 structures with largest
mean force disagreement are added to the training set.
The final C-NNP models are then obtained
after stringent training of the NNP members with
tight convergence criteria, as mentioned below in detail.

The chemical environment around each atom
is described by a set of atom--centered symmetry functions~\cite{Behler2011/10.1063/1.3553717},
which transform the structure into suitable
input for the atomic NNs.
We applied a general set of symmetry functions
to all systems studied here that can
be automatically generated for a new system of interest
within the AML package.
The structural information for angular and radial
symmetry functions is restricted to a
radial cutoff of 12\,bohr by a cosine cutoff function.
For every pair of elements we employ ten radial symmetry functions,
with fixed Gaussian width of 0.308\,bohr, which are equally distributed
within the 12\,bohr cutoff.
For every triple of elements we use four angular symmetry functions
with a fixed Gaussian width of 0.012\,bohr, $\lambda=\pm 1$ and
$\zeta$ of 1 and 4, respectively.
All symmetry function values are
scaled and centered based on the average
and range of the individual symmetry functions
encountered in the training set according to
\begin{align}
    G^{i} =  \frac{G^i - G^i_\text{avg}}{G^i_\text{max}-G^i_\text{min}}.
\end{align}

All NNs consist of two hidden layers with 20 neurons, while
the hyperbolic tangent was used as activation function
for all layers, except the output neuron, which features a
linear activation function.
NNP optimizations are performed with the open-source n2p2
code~\cite{Singraber2019/10.1021/acs.jctc.8b01092}
and the optimization parameters have
been chosen according to the detailed benchmarking
of this code for water~\cite{Singraber2019/10.1021/acs.jctc.8b01092}.
Each C-NNP model is made up of 8 NNP members, which are
constructed by random subsampling of the full reference data,
where 10\% of the points are left out in each case to
impose the required diversity between C-NNP members.
After different random initialization for each committee member,
the weights and biases of the NNs were optimized using the
parallel multistream version~\cite{Singraber2019/10.1021/acs.jctc.8b01092}
of the adaptive global extended Kalman filter
as implemented in n2p2.
C-NNPs used for QbC were optimized for 15 epochs with 6 streams,
while the final C-NNPs, to be used for simulations,
were optimized for 50 epochs with 24 streams.
All training input files, training sets
and parameters of the final models are publicly available
at \url{https://github.com/water-ice-group/simple-MLP}.

\subsection*{AIMD simulations}
\label{subsec:aimd}

All \textit{ab initio} molecular dynamics simulations used as input for our machine learning
framework haven been performed with the CP2K software package~\cite{Kuhne2020/10.1063/5.0007045}.

The fluoride ion in water was described at the
hybrid DFT level with the revPBE0 functional
and D3 dispersion correction.
The wavefunction was represented up to a plane wave 
cutoff of 400\,Ry in conjunction with the 
TZV2P basis set and GTH pseudopotentials.
The Hartree-Fock exchange calculation is speed up
using the auxiliary density matrix methods as
implemented in C2PK.
A single fluoride ion was described in a periodic
box of 64 water molecules with a cell size of 12.445\,\AA{}.
The system was propagated in the NVT ensemble at 300\,K
with a molecular dynamics timestep of 0.5\,fs.
Equilbiration with a CSVR thermostat and 30\,fs coupling constant
was performed for 3\,ps, while the temperature for the 50\,ps
production run was maintained with a CSVR thermostat and 1\,ps
coupling constant.

The sulfate ion in water was described with the
BLYP functional in combination with the D3 dispersion correction.
A plane wave cutoff of 280\,Ry was used, while the
molecularly optimized TZV2P atomic basis set was employed for all elements
in combination with GTH pseudopotentials.
The system contains a single sulfate ion and 64 water molecules
in a 12.41\,\AA{} periodic box.
This setup was simulated in the NVT ensemble with
a time step of 0.5\,fs, while the temperature was
maintained with a CSVR thermostat with a 50\,fs coupling constant.
Equilibration was performed for 5\,ps followed by a 30\,ps
production run.
This simulation has been used in a previous study
on the effects of polarization for the properties
of the sulfate ion~\cite{Pegado2012/10.1039/c2cp40711f}.

Simulations of water confined in carbon and hexagonal boron nitride nanotubes
were performed for (12,12) armchair nanotubes with a length of 3 unit cells
at a water density of 1.0\,$\mathrm{g/cm}^{3}$.
This results in 288 wall atoms and 65 water molecules for the carbon nanotube
and 68 water molecules for the hexagonal boron nitride nanotube.
The PBE functional with D3 dispersion correction was used, in combination
with GTH pseudopotentials, a 460\,Ry plane wave cutoff and the
DZVP molecularly optimised basis set.
Deuterium masses were used for the hydrogen atoms and the molecular
dynamics time step was set to 1.0\,fs.
Systems were pre-equilibrated for 5\,ps at a temperature of 500\,K
using velocity rescaling, while keeping the positions of the atoms
in the confining nanotubes fixed.
Production runs in the NVT ensemble were then performed using Langevin dynamics,
at a temperature of 330 K.
Statistics were collected for approximately 130 ps for each system.

For water confined within molybdenum disulphide,
the \textit{ab initio} reference simulation consists
of 109 water molecules confined by single layer \ch{MoS2} sheets
of 168 atoms in a 22.545, 22.314, 11.500\,\AA{} periodic box.
The optB88-vdW functional was used, with GTH pseudopotentials, a 550\,Ry plane wave cutoff,
a relative cutoff of 60\,Ry and the DZVP molecularly optimised basis set for all elements.
Equilibration was performed in the NVT ensemble, with a Nosé-Hoover chain thermostats of length 5 to maintain a temperature of 500\,K for 5\,ps.
This was followed by a second equilibration stage at 400\,K for a further 5\,ps.
Final data collection was performed at 400\,K over a 30\,ps period.

Water on the rutile titanium dioxide (110) surface was described by a system
consisting of 80 water molecules on four O-Ti-O trilayers
in a periodic box of dimension 11.836, 12.9938, 42.00\,\AA{}.
This results in a 1.5\,nm thick water film on the four
trilayers with additional 15\,\AA{} vacuum to separate the
periodic images in z-direction.
The optB88-vdW functional was used in combination with GTH pseudopotential,
a 400\,Ry plane wave cutoff and the DZVP molecularly optimised basis set for all elements.
After equilibration, the system was propagated for 30\,ps in the NVT ensemble at 300\,K
maintained by a Nosé-Hoover chain thermostat of length 4 with a coupling constant
of 40\,fs.
A 1.0\,fs timestep in combination with deuterium masses for hydrogen atoms
was used and the atoms of the lowest O-Ti-O trilayer --- not in contact with
water --- were kept fixed.

\FloatBarrier

\section*{References}

%

\end{bibunit}

\end{document}